\documentclass[twocolumn,letter]{jpsj2}

\title{$^1$H-NMR Study of the Random Bond Effect in the Quantum Spin System (CH$_3$)$_2$CHNH$_3$Cu(Cl$_x$Br$_{1-x}$)$_3$}

\author{Tomoya \textsc{Adachi}, Keishi \textsc{Kanada}, Takehiro \textsc{Saito}, Akira \textsc{Oosawa} and Takayuki \textsc{Goto}}

\inst{Department of Physics, Sophia University, 7-1 Kioi-cho, Chiyoda-ku, Tokyo 102-8554, Japan\\}

\recdate{\today}

\abst{The spin-lattice relaxation rate $T_1^{-1}$ of $^1$H-NMR has been measured in (CH$_3$)$_2$CHNH$_3$Cu(Cl$_x$Br$_{1-x}$)$_3$ with $x=0.88$, which is reportedly a gapped system with a singlet ground state from the previous macroscopic magnetization and specific heat measurements, in order to investigate the bond randomness effect microscopically in the gapped composite Haldane system (CH$_3$)$_2$CHNH$_3$CuCl$_3$. It was discovered that the spin-lattice relaxation rate $T_1^{-1}$ in the present system includes both fast and slow relaxation parts indicative of the gapless magnetic ground state and the gapped singlet ground state, respectively. We discuss the obtained results with the previous macroscopic magnetization and specific heat measurements together with the microscopic $\mu$SR experiments.}

\kword{(CH$_3$)$_2$CHNH$_3$Cu(Cl$_x$Br$_{1-x}$)$_3$, quantum spin system, spin gap, bond randomness, $^1$H-NMR, spin-lattice relaxation rate $T_1^{-1}$}

\begin{document}
\sloppy
\maketitle

The bond randomness effects in spin gap systems yield a rich variety of novel phases such as the impurity-induced magnetic ordered phase \cite{Regnault,Waki,Yasuda}, the Bose-glass phase, \cite{OosawaTlK,Shindo,Fujiwara,Suzuki,Fisher} and the random-dimer phase \cite{Hyman}. \par
The title compound (CH$_3$)$_2$CHNH$_3$Cu(Cl$_x$Br$_{1-x}$)$_3$ (abbreviated as IPACu(Cl$_x$Br$_{1-x}$)$_3$) shows the impurity-induced antiferromagnetic ordered phase. Both parent compounds IPACuCl$_3$ and IPACuBr$_3$ are the spin gap systems. The magnitude of the excitation gap $\Delta$ between the singlet ground state and the triplet excited states in IPACuCl$_3$ was estimated as 17.1$\sim$18.1 K by means of magnetic susceptibility measurements \cite{Manakasus}. From the viewpoint of the crystal structure of IPACuCl$_3$, the origin of the spin gap was expected to be the $S=\frac{1}{2}$ ferromagnetic-antiferromagnetic alternating chain along the $c$-axis \cite{Manakasus}. However, recently, it was suggested that IPACuCl$_3$ should be characterized as the spin ladder along the $a$-axis with the strongly coupled ferromagnetic rungs, namely, the antiferromagnetic chain with an effective $S=1$ "composite Haldane chain," and the excitation gap was re-estimated as 13.6 K by means of neutron inelastic scattering experiments \cite{Masuda}. Meanwhile, IPACuBr$_3$ has been characterized as the $S=\frac{1}{2}$ antiferromagnetic-antiferromagnetic alternating chain with a singlet dimer ground state and an excitation gap $\Delta= 98$ K \cite{ManakaBr}. However, IPACuBr$_3$ may also be recharacterized as the spin ladder system when the neutron inelastic scattering experiments will be carried out, in the same manner as IPACuCl$_3$. \par
The mixed system IPACu(Cl$_x$Br$_{1-x}$)$_3$ was studied by means of the magnetization and specific heat measurements, and it was reported that the bond-randomness-induced antiferromagnetic ordered phase with $T_{\rm N} = 13\sim$17 K emerges in the region $0.44 < x < 0.87$ along with the Haldane phase in $x \ge$ 0.87 and the singlet dimer phase in $x \le$ 0.44 \cite{Manakamixsus,Manakamixmag} and that the phase boundaries between these phases are of the first order. Quite recently, the present authors microscopically studied the mixed system IPACu(Cl$_x$Br$_{1-x}$)$_3$ by means of NMR \cite{Kanada,KanadaQuBS} and $\mu$SR \cite{Saito} measurements. Microscopic evidence of the impurity-induced antiferromagnetic ordered phase, such as the clear splitting of the $^1$H-NMR spectra and the clear rotation of the $\mu$SR time spectra below $T_{\rm N}$, was observed in IPACu(Cl$_x$Br$_{1-x}$)$_3$ with $x=0.85$. Further, in order to investigate the bond randomness effect in the Haldane phase of this system microscopically, we performed muon spin relaxation measurements in IPACu(Cl$_x$Br$_{1-x}$)$_3$ with $x=0.95$ and observed an anomalous enhancement of the relaxation rate indicative of the existence of magnetic fluctuations at low temperatures, in spite of the singlet Haldane phase. \par
As mentioned above, based on the previous magnetization measurements, it was concluded that the ground state is the singlet Haldane phase in $x \ge 0.87$, while from the recent $\mu$SR measurements, it was concluded that magnetic fluctuations exist in the ground state of $x=0.95$. In order to investigate these two conflicting conclusions, we measured the spin-lattice relaxation rate $T_1^{-1}$ of $^1$H-NMR in IPACu(Cl$_x$Br$_{1-x}$)$_3$ with $x=0.88$ and observed {\it both} fluctuations indicative of the gapped singlet ground state and the gapless magnetic one. In this letter, we report and discuss the results. \par
Single crystals of IPACu(Cl$_x$Br$_{1-x}$)$_3$ with $x=0.88$ were prepared by the evaporation method. An isopropanol solution of isopropylamine hydrochloride, copper(II) chloride dihydrate, isopropylamine hydrobromide, and copper (II) bromide was placed in a bowl, which was maintained at $30 \pm 0.1$ $^{\circ}$C, in an atmosphere of flowing nitrogen gas during the entire period of crystal growth, which was approximately two months. Crystals with three orthogonal surfaces were obtained. These three planes were termed A-, B-, and C-planes. The definition of these planes is given in ref. \citen{Manakasus}. The typical size of the obtained crystals was around $2 \times 3 \times 8$ mm$^3$ with a rectangular shape, as reported in a previous paper \cite{Manakamixsus}. The content of Cl, $x=0.88$, was determined using the inductively coupled plasma spectrometry for three tiny fragments chipped off from different points of the crystal. \par
The spin-lattice relaxation rate $T_1^{-1}$ of $^1$H-NMR with $\nu=118$ MHz was measured by the saturation-recovery method with a pulse train using a 4 K cryogen-free refrigerator set in a 6 T cryogen-free superconducting magnet. Relaxation curves were traced until the difference between the nuclear magnetization and its saturation value was 1 \%. There are ten inequivalent proton sites in the unit cell of the present system. Each inequivalent proton site has different distances such that it is exposed to different hyperfine fields from the nearest magnetic Cu site. Since the nuclear spin-spin interactions between inequivalent protons result in a small hyperfine field, the fine structure of $^1$H-NMR spectra produced by inequivalent protons were smeared and overlapped broad peaks were observed. In the present experiments, we measured the spin-lattice relaxation rate $T_1^{-1}$ of the overlapped broad peaks. We confirmed that there was no significant difference in $T_1^{-1}$ at any position of the spectrum. \par
\begin{figure}[t]
\begin{center}
\includegraphics[width=80mm]{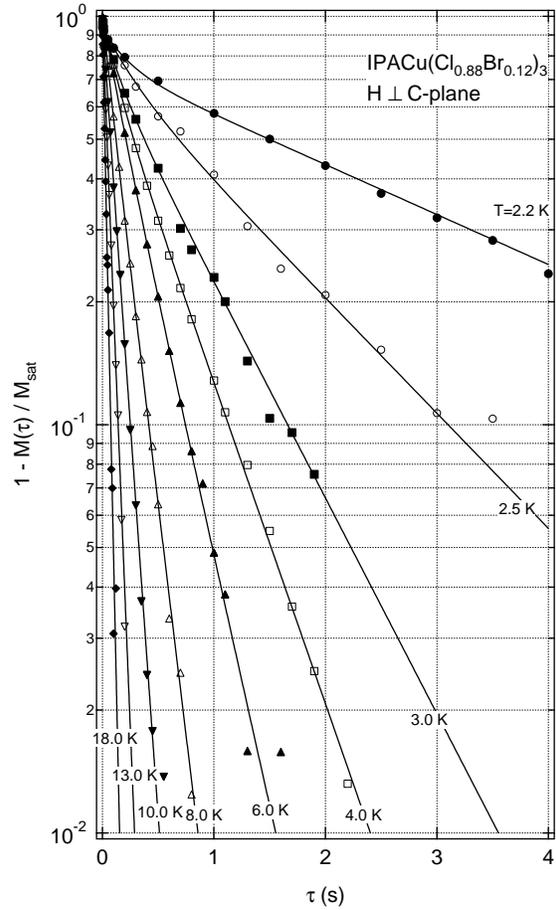}\\
\end{center}
\caption{Nuclear magnetization recovery of $^1$H-NMR at various temperatures for the $H \perp$ C-plane in IPACu(Cl$_{0.88}$Br$_{0.12}$)$_3$. Solid lines denote the results of fitting by eq. (\ref{eq1}). \label{Fig1}}
\end{figure}

Figure \ref{Fig1} shows the nuclear magnetization recovery of $^1$H-NMR at various temperatures for the $H \perp$ C-plane in IPACu(Cl$_{0.88}$Br$_{0.12}$)$_3$. As shown in Fig. \ref{Fig1}, nuclear magnetization exhibits fast recovery at high temperatures. However, with a decrease in temperature, we can clearly observe that the recovery of the nuclear magnetization becomes slower and does not obey the single exponential function. Based on the assumption that the spin-lattice relaxation includes fast and slow relaxation parts, we fitted the obtained results by the following equation

\begin{equation}
\label{eq1}
1- \frac{M ( \tau )}{M_{\rm sat}} = a \exp \left[ - \frac{\tau}{(T_1)_{\rm fast}} \right] + b \exp \left[ - \frac{\tau}{(T_1)_{\rm slow}} \right].
\end{equation}

Reasonable fits can be obtained, as shown in Fig. \ref{Fig1}, and the spin-lattice relaxation rate $T_1^{-1}$ for both parts is obtained for each temperature. In this fitting, the ratios of the two relaxation parts $a$ and $b$ are assumed to be temperature independent and are estimated to be 0.2 and 0.8, respectively. \par
\begin{figure}[t]
\begin{center}
\includegraphics[width=80mm]{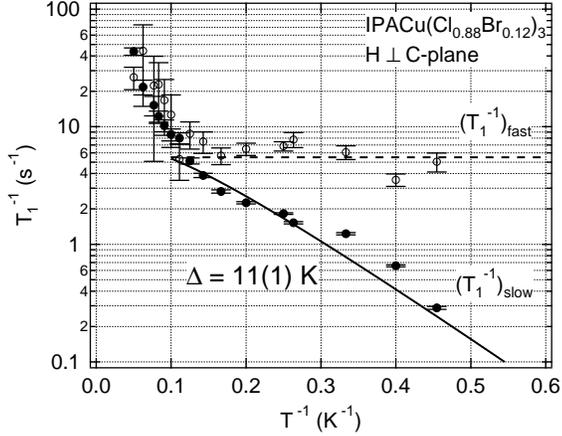}\\
\end{center}
\caption{Temperature dependence of fast and slow relaxation parts of the spin-lattice relaxation rate $T_1^{-1}$ of $^1$H-NMR for the $H \perp$ C-plane in IPACu(Cl$_{0.88}$Br$_{0.12}$)$_3$. The solid line denotes the result of fitting by eq. (\ref{eq2}) for the slow relaxation part of $T_1^{-1}$. The dashed line is the guide for the eyes. \label{Fig2}}
\end{figure}

Figure \ref{Fig2} shows the temperature dependence of the fast and slow relaxation parts of the spin-lattice relaxation rate $T_1^{-1}$ of $^1$H-NMR for the $H \perp$ C-plane in IPACu(Cl$_{0.88}$Br$_{0.12}$)$_3$. At high temperatures, both $T_1^{-1}$ show a similar decrease with decreasing temperature due to the development of antiferromagnetic correlations, as observed in the previous magnetic susceptibility measurements \cite{Manakamixsus}. However, below $T$$\sim$10 K, each $T_1^{-1}$ exhibits different behavior, namely $(T_1^{-1})_{\rm fast}$ becomes constant, while $(T_1^{-1})_{\rm slow}$ decreases more rapidly on decreasing the temperature. We fitted the temperature dependence of $(T_1^{-1})_{\rm slow}$ at a low temperature by the following equation

\begin{equation}
\label{eq2}
\frac{1}{(T_1)_{\rm slow}} \propto \frac{1}{\sqrt{T}} \exp \left( - \frac{\Delta}{k_{\rm B} T} \right),
\end{equation}

which indicates the existence of the excitation gap from the singlet ground state and was used for the estimation of the excitation gap $\Delta$ in the previous magnetic susceptibility and specific heat measurements \cite{Manakamixsus}. The excitation gap $\Delta$ was estimated as 11(1) K, as indicated by the solid line in Fig. \ref{Fig2}. \par
Herein, we discuss the obtained results. First, it is noted that there exists a correspondence between the magnitudes of $(T_1^{-1})_{\rm fast}$ and $(T_1^{-1})_{\rm slow}$ within the error bars above $T$$\sim$10 K, as shown in Fig. \ref{Fig2}; therefore, the nuclear magnetization recovery observed above $T$$\sim$10 K can also be fitted by the single exponential function. In the present experiment, we discovered that nuclear magnetization recovery below $T$$\sim$10 K cannot be expressed as a single exponential function, that the spin-lattice relaxation rate $T_1^{-1}$ consists of fast and slow relaxation parts, and that the slow part of $T_1^{-1}$ at low temperatures exhibits gapped behavior with $\Delta=11(1)$ K in IPACu(Cl$_{0.88}$Br$_{0.12}$)$_3$. As mentioned above, it has been reported that the ground state of IPACu(Cl$_{0.88}$Br$_{0.12}$)$_3$ is the singlet Haldane phase with an excitation gap $\Delta$$\sim$12 K, which is almost the same as the parent compound IPACuCl$_3$ in the previous magnetization and specific heat measurements \cite{Manakamixsus,Manakamixmag}. Hence, we can expect that the gapped behavior of the slow relaxation part of $T_1^{-1}$ indicates the existence of the gapped singlet ground state, as observed in the macroscopic magnetization and specific heat measurements \cite{Manakamixsus,Manakamixmag}. In addition, the fast relaxation part of $T_1^{-1}$ was also observed in the present experiments. Since the fast relaxation part of $T_1^{-1}$ becomes constant with a decrease in temperature, we can expect that the fast part of $T_1^{-1}$ indicates that there is the magnetic ground state composed of localized moments, in contrast with the slow relaxation part. In the previous $\mu$SR experiments for IPACu(Cl$_{0.95}$Br$_{0.05}$)$_3$ \cite{Saito}, muon spin relaxation with two components was observed at $T=0.33$ K, as observed in the present NMR experiments for IPACu(Cl$_{0.88}$Br$_{0.12}$)$_3$; further, it was concluded that one of the two components indicates the existence of the magnetic instability in the ground state. Hence, we can expect that the fast part of the spin-lattice relaxation rate $(T_1^{-1})_{\rm fast}$ observed in the present NMR experiments corresponds to the observed muon spin relaxation indicating the magnetic instability and also supports the existence of magnetic fluctuations in the ground state. The other of two components observed in the $\mu$SR experiments was also concluded to originate due to the quasistatic nuclear spin part, that is, a part of the muons was unaffected by the magnetic fluctuation of electronic moment at the Cu sites. We infer that this component observed in the $\mu$SR experiments reflects the gapped singlet ground state, as observed in the present NMR experiments. \par
That the two spin-lattice relaxation components originate due to the macroscopic phase separation can be ruled out on the basis of the following discussions. For one, no diverging behavior indicative of magnetic ordering above $T=2$ K was observed in $(T_1^{-1})_{\rm fast}$. It has been reported that the magnetic ground state only exists in the impurity-induced magnetic ordered phase with $T_{\rm N} = 13$$\sim$17 K for the region $0.44 < x < 0.87$ in IPACu(Cl$_x$Br$_{1-x}$)$_3$ \cite{Manakamixsus,Manakamixmag} and that the diverging behavior of the spin-lattice relaxation rate $T_1^{-1}$ of $^1$H-NMR indicative of the impurity-induced magnetic ordering was actually observed in IPACu(Cl$_x$Br$_{1-x}$)$_3$ with $x=0.85$ at $T_{\rm N} = 13.5$ K \cite{Kanada}; therefore, we can safely conclude that the observed $(T_1^{-1})_{\rm fast}$ does not reflect the magnetic ordered phase of $0.44 < x < 0.87$. Second, the changes in the temperature dependence of both $(T_1^{-1})_{\rm slow}$ and $(T_1^{-1})_{\rm fast}$ at $T$$\sim$10 K occur simultaneously and therefore we expect that the changes are cooperative. Since it has been observed that the singlet formation commences at $T$$\sim$10 K in the parent compound IPACuCl$_3$ \cite{ManakaESR1,ManakaESR2}, we expect that the change in $(T_1^{-1})_{\rm fast}$ is also associated with the singlet formation together with the exponential decrease in $(T_1^{-1})_{\rm slow}$, as shown in Fig. \ref{Fig2}. \par
From the observation of $(T_1^{-1})_{\rm slow}$ and $(T_1^{-1})_{\rm fast}$, it can be considered that there are more than two Cu sites with inequivalent magnetic properties because the spin-lattice relaxation process of $^1$H-NMR is caused by the fluctuation of the magnetic moment around the protons. It should be noted that the observed fast and slow parts do not correspond to the relaxation processes of the inequivalent protons in the unit cell of the present system because the spin-lattice relaxation rates of inequivalent protons should exhibit the same temperature dependence and should be scaled by the distance between each proton and the magnetic moment at the Cu site. Non-single exponential behavior of spin-lattice relaxation indicative of the existence of some inequivalent magnetic sites has also been observed in the doped spin-Peierls systems (Cu$_{1-x}$Mg$_x$)GeO$_3$ \cite{Itoh} and Cu(Ge$_{1-x}$Si$_x$)O$_3$ \cite{Kikuchi}. This non-single exponential behavior has been expressed in the spatially nonuniform relaxation processes, as expressed by the stretched exponential form $1- M ( \tau )/M_{\rm sat} \propto \exp [ -t / T_1 - (t/\tau_1)^{1/2} ]$, indicating that there are additional dilute magnetic moments in the homogeneous host. We tried to fit the present results using the above stretched exponential form; however, we were unable to obtain suitable results. Hence, we conclude that each magnetic moment is not isolated dilutely, but forms islands in the singlet sea in the present system. \par
From the above discussion, we expect that the ground state of the present system can be expressed as a microscopic mixture of singlets and localized magnetic moment islands. The singlets correlate to the magnetic moment islands because the Curie-law behavior indicative of the existence of paramagnetic moments has not been reported in the magnetic susceptibility measurements \cite{Manakamixsus}, and simultaneous changes in the spin-lattice relaxation rate below $T$$\sim$10 K were observed, as mentioned above. Further, it was found that the spin-lattice relaxation behavior of the present system is considerably different from that of the parent compound IPACuCl$_3$, in which the nuclear magnetization recovery can be expressed as a single exponential function, and the spin-lattice relaxation rate $T_1^{-1}$ exhibits an exponential decrease indicative of the gapped singlet ground state \cite{Kanada}. Hence, the present result indicates that the ground state of the present system is a new phase that is different from both the singlet Haldane phase of IPACuCl$_3$, as opposed to the conclusion of the previous macroscopic experiments \cite{Manakamixsus,Manakamixmag}, and the impurity-induced magnetic ordered phase in the region $0.44 < x < 0.87$ of IPACu(Cl$_x$Br$_{1-x}$)$_3$. Spin-lattice relaxation measurements at lower temperatures are required in order to investigate the ground state of the present system in a more detailed manner, especially to investigate whether the fast part of the spin-lattice relaxation rate $(T_1^{-1})_{\rm fast}$ exhibits diverging behavior or remains constant, which indicate the magnetic ordering and precursor phenomenon of the Bose-glass phase at $T=0$ \cite{Suzuki,Fujiwara}, respectively. This problem will be considered in future studies. \par
The observed nontrivial ground state composed of both the magnetic and nonmagnetic Cu sites is induced by the bond randomness effect for the gapped singlet Haldane phase of the parent compound IPACuCl$_3$ and is revealed for the first time from the observation of both gapless magnetic and gapped nonmagnetic fluctuations by the present $^1$H-NMR experiments. \par     
In conclusion, we have presented the results of the spin-lattice relaxation rate $T_1^{-1}$ measurements of $^1$H-NMR in IPACu(Cl$_x$Br$_{1-x}$)$_3$ with $x=0.88$. It was observed that the nuclear magnetization recovery of $^1$H-NMR exhibits non-single exponential behavior and can be well expressed by the model that includes the fast and slow spin-lattice relaxation processes, as shown in Fig. \ref{Fig1}. It was also found that the obtained temperature dependence of the fast and slow parts of the spin-lattice relaxation rate $T_1^{-1}$ of $^1$H-NMR exhibit the simultaneous changes below $T$$\sim$10 K, that is, the fast part $(T_1^{-1})_{\rm fast}$ becomes constant indicative of the magnetic ground state composed of localized moments, while the slow part $(T_1^{-1})_{\rm slow}$ decreases exponentially with $\Delta=11(1)$ K indicative of the gapped singlet ground state, as shown in Fig. \ref{Fig2}. From the discussions on the previous macroscopic magnetization and specific heat measurements and the microscopic $\mu$SR experiments, we expect that the ground state of the present system is a new phase in which the singlets and localized magnetic moment islands are mixed microscopically and both the singlets and magnetic moment islands are correlated with one another; this new phase is different from both the singlet Haldane phase of IPACuCl$_3$ and the impurity-induced magnetic ordered phase in the region $0.44 < x < 0.87$ of IPACu(Cl$_x$Br$_{1-x}$)$_3$. \par
We acknowledge H. Manaka and T. Suzuki for their useful suggestions in discussions. This work was supported by Grants-in-Aid for Scientific Research on Priority Areas "High Field Spin Science in 100 T" from the Ministry of Education, Science, Sports and Culture of Japan, the Saneyoshi Scholarship Foundation, and the Kurata Memorial Hitachi Science and Technology Foundation. \par

\end{document}